\documentclass[aps, preprint,prf,longbibliography]{revtex4-2}
\usepackage{dcolumn}
\usepackage{epsfig, graphics, amssymb, amsmath, color, bm}
\usepackage{epstopdf, stmaryrd}
\usepackage{amsmath}
\usepackage{color}
\usepackage{siunitx}
\usepackage{float}
\usepackage{soul}
\usepackage{pifont,wasysym}
\usepackage{color,amsmath}
\usepackage{mathrsfs}

\def\boldsymbol{\bm}

\def \sgn{\text{sgn}}

\usepackage[unicode=true,
  linktocpage,
  linkbordercolor={0.5 0.5 1},
  citebordercolor={0.5 1 0.5},
  linkcolor=black, citecolor = black, colorlinks=true]{hyperref}

\definecolor{dgreen}{rgb}{0.0, 0.4, 0.0}

\begin{document}

\title{Collinear Swimming of a Squirmer Pair in Newtonian and Shear-Thinning Fluids}

\author{Chih-Tang Liao\textsuperscript{1,2}}
\author{Ali Gürbüz\textsuperscript{1,3}}
\author{Victor Bueno Garcia\textsuperscript{1}}
\author{Yuan-Nan Young\textsuperscript{4}}
\author{Devanayagam Palaniappan\textsuperscript{5}}
\author{On Shun Pak\textsuperscript{1}\footnote{Email address for correspondence: \texttt{opak@scu.edu}}}

\affiliation{\textsuperscript{1}Department of Mechanical Engineering, Santa Clara University, Santa Clara, CA 95053, USA}
\affiliation{\textsuperscript{2}Department of Mechanical Sciences and Engineering, University of Illinois Urbana-Champaign, Urbana, Illinois, 61801, USA}
\affiliation{\textsuperscript{3} Department of Mathematics, Towson University, Towson, Maryland, 21252, USA}
\affiliation{\textsuperscript{4}Department of Mathematical Sciences, New Jersey Institute of Technology, Newark, New Jersey, 07102, USA}
\affiliation{\textsuperscript{5}Department of Mathematics and Statistics, Texas A\&M University–Corpus Christi, Corpus Christi, TX 78412, USA}

\begin{abstract}
Pairwise hydrodynamic interactions of microswimmers form the fundamental building blocks for understanding their more complex collective behaviors. In this work, we revisit the canonical problem of two interacting squirmers swimming along their common line of centers in both Newtonian and shear-thinning fluids. For the Newtonian case, we first derive an exact, closed-form solution for the axisymmetric Stokes flow generated by the interacting pair, thereby complementing prior analyses based on the reciprocal theorem approach by providing direct access to the detailed knowledge of the flow around the swimmers. The analytical solution is then used to cross-validate numerical simulations based on the finite element method. The combined theoretical and numerical investigation reveals co-swimming configurations in which the two squirmers develop identical velocities over a range of separations. We rationalize these behaviors through symmetry arguments and quantify their propulsion performance in terms of the speed and energetic cost of swimming. Furthermore, motivated by the prevalence of shear-thinning biological fluids encountered by microswimmers, we examine how this ubiquitous non-Newtonian rheological behavior modifies the propulsion characteristics of these co-swimming pairs. Taken together, our results establish quantitative benchmarks for interacting squirmers in both Newtonian and shear-thinning fluids, laying the groundwork for future studies of many-body dynamics of microswimmers in complex fluid environments.
\end{abstract}

\maketitle

\section{Introduction}\label{sec:Intro}
Locomotion at the microscale has attracted significant interest due to the fundamental challenge of swimming in a regime where inertia is negligible \cite{purcell1977life, Fauci2006, lauga2009}, as well as its potential biomedical and environmental applications, including targeted drug delivery, microsurgery, and environmental remediation \cite{wu2020, yan2022, Urso2023}. These fundamental and practical interests have engendered extensive theoretical, computational, and experimental studies aimed at elucidating individual swimming behaviors and hydrodynamic interactions between microswimmers in complex fluid environments  \cite{lauga2009, Li2021, Gompper_2020, Ju2025}. In particular, a canonical theoretical framework for microscale propulsion is the squirmer model, which was first considered by Lighthill \cite{lighthill1952} and Blake \cite{blake1971} for studying ciliary propulsion. In this model, the beating of cilia is represented by surface velocity distributions on the swimmer boundary, which drive fluid motion and generate self-propulsion at low Reynolds numbers. In addition to modeling swimming of ciliates such as \textit{Paramecium caudatum} \cite{Ishikawa2006JEB} and \textit{Volvox carteri} \cite{Short2006}, the squirmer model has become a widely used general model for low-Reynolds-number locomotion \cite{Pedley16}, as it captures the essential hydrodynamic signatures of a swimmer through a systematic decomposition of the flow field into hydrodynamic singularities. In particular, the first squirming mode corresponds to a source dipole and sets the swimming speed of an isolated swimmer, while the second mode corresponds to a force dipole (stresslet) that determines whether the swimmer behaves as a pusher (e.g., \textit{Escherichia coli}), puller (e.g., \textit{Chlamydomonas}), or a neutral squirmer. Owing to its simplicity and versatility, the model has been adopted to investigate a wide range of phenomena in low–Reynolds-number locomotion, including swimming under geometric confinement, the influence of complex rheology, collective dynamics, and active suspensions \cite{Ishikawa2024}.

Pairwise interactions form the basis for understanding more complex many-body interactions. Considerable efforts have been devoted to elucidating the hydrodynamics of an interacting squirmer pair. In particular, Ishikawa \textit{et al.} \cite{Ishikawa2006} developed a database for an interacting pair of squirmers, where near- and far-field interactions are treated analytically using lubrication theory and Faxén relations, respectively, while the intermediate-separation regime is resolved numerically via the boundary element method. Lubrication theory has also been shown to effectively capture the scattering dynamics of squirmers \cite{Darveniza2022}. Another line of inquiry considers assemblies of microswimmers, where interactions among swimmers give rise to emergent behaviors that may be exploited for enhanced transport, mixing, and other microscale manipulations \cite{Wang2015, Winkler2020}. In particular, prior studies have investigated dumbbell squirmers, in which individual squirmers are mechanically coupled via rigid rods or elastic springs \cite{Ishikawa2019, Clops2020, Clops2022, Ouyang2022a, Ouyang2022b}, providing insights into the design principles of microswimmer assemblies. Beyond spherical squirmers in Stokes flow, the effects of non-spherical geometry \cite{Kyoya2015, Theers2016}, inertia \cite{Li2016}, and density stratification \cite{More2021} on the hydrodynamic interactions of squirmer have also been characterized under different configurations \cite{Ishikawa2024}.

Here, we revisit the problem of two interacting squirmers to complement previous analyses in Stokes flow and to extend the investigation to non-Newtonian rheology. In a Newtonian fluid, Papavassiliou and Alexander \cite{Papavassiliou2017} employed the Lorentz reciprocal theorem, together with known Stokes flow solutions, to derive exact expressions for the swimming velocities of two interacting squirmers, thereby bypassing the need for explicit calculation of the surrounding flow field. This reciprocal theorem approach, sometimes described as \textit{``getting something from nothing"} \cite{Masoud_Stone_2019}, is particularly powerful for extracting integral relations and global properties of locomotion problems, including forces, torques, and swimming velocities \cite{stone_reciprocal, Lauga2014}. However, while the reciprocal theorem approach yields elegant and compact results for key locomotion quantities, it does not provide access to the detailed structure of the flow field itself. Here, we address this gap by explicitly calculating the full flow field generated by two interacting squirmers, thereby complementing prior analyses and enabling detailed examination of local flow features, as well as establishing benchmarks for validation of numerical simulations. Classical foundations for such calculations date back to the work of Stimson and Jeffery \cite{Stimson1926}, who obtained the exact solution for the translation of two spheres using the Stokes streamfunction in bispherical coordinates. More recently, Jabbarzadeh and Fu \cite{Jabbarzadeh2018} extended the solution to study the interaction between a squirmer and a passive spherical particle, elucidating the viscous constraints on the approach of microorganisms to target particles. Another related extension by Shum \textit{et al.} \cite{Shum_Palaniappan_Young_2025} examined the motion of a sedimenting squirmer moving normal to a planar wall. In this work, we follow a similar approach to obtain an exact, closed-form solution for the axisymmetric Stokes flow around two interacting squirmers. This exact solution is then employed to validate numerical simulations based on the finite element method. The combined theoretical and numerical investigation reveals co-swimming configurations where the two squirmers develop identical velocities across all separation distances, and we rationalize these results through symmetry considerations.

Furthermore, biological and artificial microswimmers often encounter complex biological fluids such as blood and mucus, which exhibit pronounced shear-thinning behavior \cite{hwang1969rheological, baskurt2003blood}. In these shear-thinning fluids, the shear-rate–dependent viscosity decreases with increasing local strain rate, introducing a nonlinear and spatially varying fluid response to the motion of a swimmer. While recent studies have begun to elucidate how shear-thinning rheology affects the propulsion of an isolated swimmer \cite{montenegro2013physics, velez2013waving, gagnon2014undulatory, li2015undulatory, datt2015squirming, datt2017active, qu2020effects, van2022effect}, its impact on interacting swimmers remains largely unexplored. Shear-thinning effects may be particularly significant for interacting swimmers, whose combined flow fields can generate localized regions of elevated shear, thereby substantially altering the surrounding viscosity distribution and hence the locomotion performance. As a first step, we extend our validated numerical framework to investigate how this ubiquitous non-Newtonian rheology alters the propulsion characteristics of co-swimming squirmer pairs identified in the Newtonian limit, examining both their speed and the energetic cost of swimming.

This paper is organized as follows. In \S\ref{sec:TheoreticalAnalysis}, we formulate the theoretical analysis and derive an exact, closed-form solution for the axisymmetric Stokes flow around a pair of squirmers. In \S\ref{sec:Numerical}, we use this solution to validate our numerical framework and subsequently extend the simulations to examine the effects of shear-thinning rheology. In \S\ref{sec:ResultsDiscussion}, we discuss the propulsion characteristics of a squirmer pair in both Newtonian (\S\ref{sec:Newtonian}) and shear-thinning (\S\ref{sec:Shear-thinning}) fluids, before concluding with closing remarks and perspectives on future research directions in \S\ref{sec:Conclusions}.

\begin{figure}[t]
    \centering\includegraphics[width=0.72\textwidth]{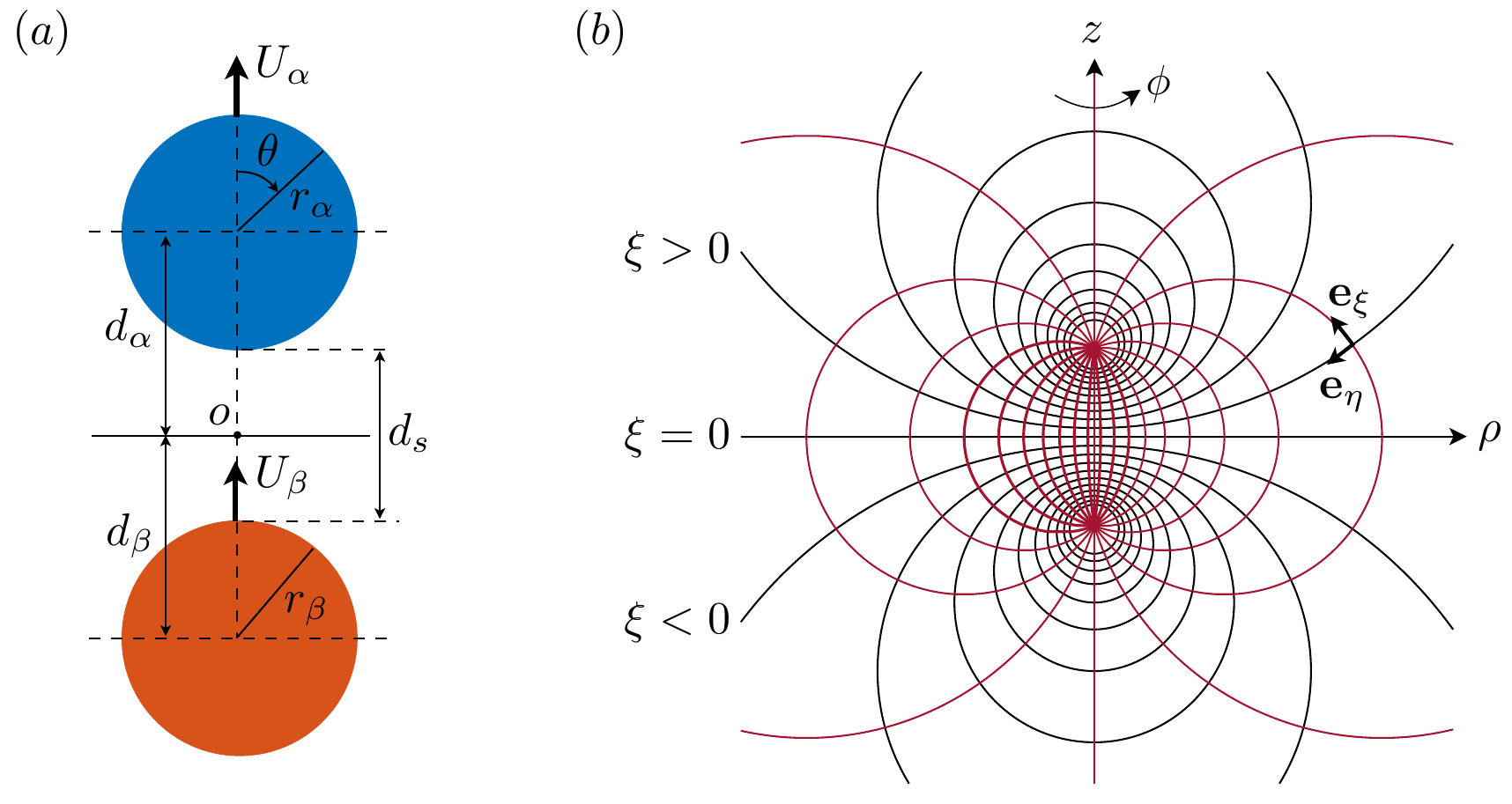}
    \caption{(a) Schematic and notation of the problem setup, where two collinear squirmers are shown. The upper squirmer, with radius $r_\alpha$, is located at a distance $d_\alpha$ above the origin ($o$) and swims with speed $U_\alpha$. The lower squirmer, with radius $r_\beta$, is located at a distance $d_\beta$ below the origin and swims with velocity $U_\beta$. The surface-to-surface separation between the two squirmers is denoted by $d_s$. (b) Relation between the bispherical coordinates $(\xi,\eta,\phi)$ and the cylindrical coordinates 
    $(\rho,\phi,z)$. The black and red lines denote $\xi-$ and $\eta-$isosurfaces, 
    respectively, and $\mathbf{e}_{\xi}$ and $\mathbf{e}_{\eta}$ represent the unit vectors.}
    \label{fig_1}
\end{figure}

\section{Theoretical Analysis} \label{sec:TheoreticalAnalysis}

\subsection{Problem setup}
We consider the motion of two spherical squirmers, labeled $\alpha$ and $\beta$, of radii $r_\alpha$ and $r_\beta$, as shown in Fig.~\ref{fig_1}(a). Each squirmer is characterized by a prescribed tangential surface velocity distribution along the polar angle $\theta$ \cite{lighthill1952,blake1971}, expressed as a series of modes corresponding to different flow singularities \cite{On_shun_general}. Locomotion studies typically retain only the first two modes \cite{Pedley16,Ishikawa2024}, 
\begin{align}
    u^{\alpha,\beta}_\theta = B_{1}^{\alpha,\beta} \sin{\theta} + \frac{1}{2}B_{2}^{\alpha,\beta}\sin{2\theta}. \label{eqn:TwoModes}
\end{align}
The first mode ($B_1^{\alpha,\beta}$) corresponds to a source dipole and sets the swimming velocity of an isolated squirmer in a Newtonian fluid \cite{lighthill1952, blake1971,Pedley16,Ishikawa2024}, given by $U^\infty_{\alpha,\beta} = 2B^{\alpha, \beta}_1/3$ . The second mode ($B_2^{\alpha,\beta}$) corresponds to a force-dipole and determines the swimming characteristics of the squirmer. Specifically, the sign of $B^{\alpha,\beta}_2$ determines whether a squirmer behaves as a puller ($B^{\alpha,\beta}_2>0$; \textit{e.g.}, \textit{Chlamydomonas}), a pusher ($B^{\alpha,\beta}_2<0$; \textit{e.g.}, \textit{Escheriochia coli}), or a neutral squirmer ($B^{\alpha,\beta}_2=0$). In this work, we also focus on the first two modes of tangential surface velocity distribution and determine the resulting swimming velocities of the squirmers, $U_\alpha$ and $U_\beta$, along their line of centers in an unbounded fluid domain.

The momentum and continuity equations for an incompressible flow in the low-Reynolds-number limit are, respectively, given by 
\begin{align}
\nabla \cdot \boldsymbol{\sigma} = \boldsymbol{0}, \quad \nabla \cdot \mathbf{u} = 0, \label{eqn:MomentumContinuity}
\end{align}
where $\mathbf{u}$ is the fluid velocity, $\boldsymbol{\sigma}=-p \mathbf{I}+\boldsymbol{\tau}$ is the stress tensor, and $p$ and $\boldsymbol{\tau}$ are the pressure and deviatoric stress, respectively. For a Newtonian fluid, the constitutive equation for the deviatoric stress is given by $\boldsymbol{\tau} = \mu_0 \dot{\boldsymbol{\gamma}}$, where $\mu_0$ is the constant dynamic viscosity and $\dot{\boldsymbol{\gamma}} = \nabla \mathbf{u}+(\nabla \mathbf{u})^T$ is the strain-rate tensor, resulting in the Stokes equation
\begin{align}
\mu_0 \nabla^2 \mathbf{u}  = \nabla p, \quad \nabla \cdot \mathbf{u} = 0. \label{eqn:StokesEqn}
\end{align}
The problem under consideration is axisymmetric, and we use the bispherical coordinates $(\xi,\eta,\phi)$ defined by 
\begin{equation} 
z+i\rho = ic\cot\frac{1}{2}(\eta+i\xi),
\label{eq_2}
\end{equation}
where $c>0$ is a constant, and $\rho$ and $z$ are the cylindrical coordinates [Fig.~\ref{fig_1}(b)], or equivalently,
\begin{equation} 
\rho = \frac{c\sin\eta}{\cosh\xi-\cos\eta}, \quad z = \frac{c\sinh\xi}{\cosh\xi-\cos\eta} \cdot
\label{eq_3}
\end{equation}
In the present application, it is only necessary to consider the situation $\rho>0$, which corresponds to $-\infty<\xi<\infty$ and $0\le\eta\le\pi$. The inverse relation is given by
\begin{equation}
\xi+i\eta = \log\frac{\rho+i(z+c)}{\rho+i(z-c)} \cdot
\label{eq_4}
\end{equation}
The surfaces of constant $\xi$ correspond to non-concentric spheres whose centers lie along the $z$--axis at a distance $d = c \coth\xi$ from the origin. The surfaces of the two spherical squirmers are therefore given by $\xi=\alpha$ and $\xi=\beta$, respectively. We choose $\alpha$, $\beta$, and $c$ so that squirmer $\alpha$ has radius $r_\alpha$ with its center located a distance $d_\alpha$ above the origin, while squirmer $\beta$ has radius $r_\beta$ with its center a distance $d_\beta$ below the origin: 
\begin{align}
r_\alpha =\frac{c}{\sinh\alpha}, \quad r_\beta = -\frac{c}{\sinh\beta} \nonumber, \\ 
d_\alpha = c\coth\alpha, \quad d_\beta = -c\coth\beta.
\end{align}
The separation between the two squirmers is then $d_\text{s}=d_\alpha+d_\beta-(r_\alpha+r_\beta)$. In what follows, we focus on the symmetric configuration of two equal-sized squirmers by setting $\beta = -\alpha$ and $r_\alpha = r_\beta = a$.

\subsection{Method of solution}
For axisymmetric flows, the velocity field may be expressed as the curl of a vector potential, ${\bm u} = \nabla\times \bigg[\psi(\xi,\eta){\bm e}_{\phi}\bigg]$, where $\boldsymbol{e}_\phi$ is the azimuthal unit vector and $\psi(\xi,\eta)$ is the Stokes streamfunction. Under this representation, the Stokes equation, Eq.~(\ref{eqn:StokesEqn}), becomes
\begin{equation} \label{eq_2.7}
D^4\psi = 0,
\end{equation} where the differential operator $D^2$ in bispherical coordinates is given by
\begin{equation} \label{eq_5}
D^2 \equiv \frac{\cosh\xi-\mu}{c^2}\left\{\frac{\partial}{\partial \xi}\left[(\cosh\xi-\mu)\frac{\partial}{\partial\xi}\right]+(1-\mu^2)\frac{\partial}{\partial \mu}\left[(\cosh \xi-\mu)\frac{\partial}{\partial\mu}\right]\right\},
\end{equation}
with $\mu=\cos\eta$. 

A general solution to Eq.~(\ref{eq_2.7}) for the streamfunction formulated by Stimson and Jeffery \cite{Stimson1926} takes the following form 
\begin{equation} \label{eq_psi}
\psi(\xi,\eta) = \left(\cosh\xi-\mu\right)^{-3/2} \sum^{\infty}_{n=1} \chi_n\left(\xi\right)V_n\left(\mu\right),
\end{equation}where 
\begin{align} \label{eq_chi}
\chi_n(\xi) &= \mathcal{A}_n \cosh\left[\left(n-\frac{1}{2}\right)\xi\right] + \mathcal{B}_n \sinh\left[\left(n-\frac{1}{2}\right)\xi \right] \nonumber \\
&+ \mathcal{C}_n \cosh\left[\left(n+\frac{3}{2}\right)\xi\right] + \mathcal{D}_n \sinh\left[\left(n+\frac{3}{2}\right)\xi\right],
\end{align} 
and 
\begin{align}
 V_n(\mu) = P_{n-1}(\mu) - P_{n+1}(\mu).   
\end{align}
The function $V_n(\mu)$ satisfies  
\begin{equation} 
 (1-\mu^2)V_n^{''} + n(n+1)V_n = 0.
\end{equation}
We now determine the unknown coefficients in Eq.~(\ref{eq_chi}) by expressing the prescribed slip velocities on the squirmer surface in Eq.~(\ref{eqn:TwoModes}) in bispherical coordinates as infinite series involving the basis function $V_n (\mu)$. Together with the impenetrability condition, these boundary conditions now yield an algebraic system of equations for the unknown coefficients similar to that for the drag problem in Stimson and Jeffery \cite{Stimson1926}. This approach circumvents the need to deal with difference equations \cite{Jabbarzadeh2018}, leading to explicit solutions for the algebraic system. For equal spheres ($\beta = -\alpha$), these coefficients are explicitly given by
\begin{subequations}
\begin{align}
\label{eq_An}
\mathcal{A}_n &= \frac{k}{\Delta_1}(U_\alpha+U_\beta)(2n+3)\bigg[(2n-1)-(2n+1)e^{2\alpha}+2e^{-(2n+1)\alpha}\bigg] \nonumber \\
&+ \frac{2k}{\Delta_1}(B_1^\alpha+B_1^{\beta})(2n-1)(2n+3)\bigg[(e^{2\alpha}-1) + (1-e^{-2\alpha})e^{-(2n+1)\alpha}\bigg] \nonumber \\
&+\frac{16}{3}\frac{k}{\Delta_1}(B_2^{\alpha}-B_2^\beta)\cosh\bigg((n+\frac{3}{2})\alpha\bigg) M_2, \\
\label{eq_Bn}
\mathcal{B}_n &= \frac{k}{\Delta_2}(U_\alpha-U_\beta)(2n+3)\bigg[(2n-1)-(2n+1)e^{2\alpha}-2e^{-(2n+1)\alpha}\bigg] \nonumber \\
&+ \frac{2k}{\Delta_2}(B_1^\beta-B_1^{\alpha})(2n-1)(2n+3)\bigg[(1-e^{2\alpha}) + (1-e^{-2\alpha})e^{-(2n+1)\alpha}\bigg] \nonumber \\
&+\frac{16}{3}\frac{k}{\Delta_2}(B_2^{\alpha}+B_2^\beta)\sinh\bigg((n+\frac{3}{2})\alpha\bigg) M_2, \\
\label{eq_Cn}
\mathcal{C}_n &= \frac{k}{\Delta_1}(U_\alpha+U_\beta)(2n-1)\bigg[(2n+3)-(2n+1)e^{-2\alpha}+2e^{-(2n+1)\alpha}\bigg] \nonumber \\
&- \frac{2k}{\Delta_1}(B_1^\alpha+B_1^{\beta})(2n-1)(2n+3)\bigg[(1-e^{-2\alpha}) +(e^{2\alpha}-1)e^{-(2n+1)\alpha}\bigg] \nonumber \\
&-\frac{16}{3}\frac{k}{\Delta_1}(B_2^{\alpha}-B_2^\beta)\cosh\bigg((n-\frac{1}{2})\alpha\bigg) M_2, \\
\label{eq_Dn}
\mathcal{D}_n &= \frac{k}{\Delta_2}(U_\alpha-U_\beta)(2n-1)\bigg[(2n+3)-(2n+1)e^{-2\alpha}-2e^{-(2n+1)\alpha}\bigg] \nonumber \\
&+ \frac{2k}{\Delta_2}(B_1^\beta-B_1^{\alpha})(2n-1)(2n+3)\bigg[(1-e^{-2\alpha}) + (1-e^{2\alpha})e^{-(2n+1)\alpha}\bigg] \nonumber \\
&-\frac{16}{3}\frac{k}{\Delta_2}(B_2^{\alpha}+B_2^\beta)\sinh\bigg((n-\frac{1}{2})\alpha\bigg) M_2, 
\end{align}
\end{subequations}
where 
\begin{subequations}
\begin{align}
\label{eq_M2}
%k
k &= \frac{c^2n(n+1)}{\sqrt{2}(2n-1)(2n+1)(2n+3)}, \\
%\Delta_1
\Delta_1 &= 2\bigg[2\sinh\bigg((2n+1)\alpha\bigg)+(2n+1)\sinh2\alpha\bigg], \\
%$\Delta_2$
\Delta_2 &= 2\bigg[2\sinh\bigg((2n+1)\alpha\bigg)-(2n+1)\sinh2\alpha\bigg], \\
%M_2(n,\alpha)
M_2 &= \sinh\alpha\bigg\{(n-1)(n+2)\bigg[(2n+3)e^{-(n-\frac{1}{2})\alpha} - (2n-1)e^{-(n+\frac{3}{2})\alpha}\bigg]\bigg\} \nonumber \\
&- \frac{5}{4}\bigg[(n-1)(2n+3)e^{-(n-\frac{3}{2})\alpha} + (2n+1)e^{-(n+\frac{1}{2})\alpha} \nonumber \\ 
&- (n+2)(2n-1)e^{-(n+\frac{5}{2})\alpha}\bigg]. 
\end{align}
\end{subequations}

\subsection{Swimming velocity}
The hydrodynamic force acting on each sphere is given by Stimson and Jeffery \cite{Stimson1926} as
\begin{equation}
F_{\alpha, \beta} = -\frac{\pi\mu_0 2\sqrt{2}}{c}\sum_{n=1}^{\infty}(2n+1)\bigg(\mathcal{A}_n\pm \mathcal{B}_n + \mathcal{C}_n \pm \mathcal{D}_n\bigg),
\end{equation}
where the plus sign is for squirmer $\alpha$ and the minus sign is for squirmer $\beta$. Enforcing the force-free condition, the swimming speeds of the two squirmers, $U_\alpha$ and $U_\beta$, read
\begin{align}
\label{eq_U1}
U_\alpha &= \frac{1}{2} \Bigg[ \bigg(\frac{\mathbb{M}_1^\text{BD}}{\lambda_\text{M}}-\frac{{\mathbb{M}_1^\text{AC}}}{\lambda_\text{SJ}}\bigg)B_1^{\alpha} - \bigg(\frac{\mathbb{M}_1^\text{BD}}{\lambda_\text{M}}+\frac{\mathbb{M}_1^\text{AC}}{\lambda_\text{SJ}}\bigg)B_1^\beta \nonumber \\&- \bigg(\frac{\mathbb{M}_2^\text{BD}}{\lambda_\text{M}}-\frac{\mathbb{M}_2^\text{AC}}{\lambda_\text{SJ}}\bigg)B_2^{\alpha} - \bigg(\frac{\mathbb{M}_2^\text{BD}}{\lambda_\text{M}}+\frac{\mathbb{M}_2^\text{AC}}{\lambda_\text{SJ}}\bigg)B_2^\beta \Bigg], 
\end{align}
and 
\begin{align}
\label{eq_U2}
U_\beta &= \frac{1}{2}  \Bigg[ \bigg(\frac{\mathbb{M}_1^\text{BD}}{\lambda_\text{M}}-\frac{\mathbb{M}_1^\text{AC}}{\lambda_\text{SJ}}\bigg)B_1^\beta - \bigg(\frac{\mathbb{M}_1^\text{BD}}{\lambda_\text{M}}+\frac{\mathbb{M}_1^\text{AC}}{\lambda_\text{SJ}}\bigg)B_1^{\alpha} \nonumber \\&+ \bigg(\frac{\mathbb{M}_2^\text{BD}}{\lambda_\text{M}}+\frac{\mathbb{M}_2^\text{AC}}{\lambda_\text{SJ}}\bigg)B_2^{\alpha} + \bigg(\frac{\mathbb{M}_2^\text{BD}}{\lambda_\text{M}}-\frac{\mathbb{M}_2^\text{AC}}{\lambda_\text{SJ}}\bigg)B_2^\beta \Bigg],
\end{align}
where
\begin{equation} \label{eq_l_sjs}
\lambda_\text{SJ} = -\sinh\alpha\sum_{n=1}^{\infty}\frac{2n(n+1)}{(2n-1)(2n+3)}\bigg[1 - \frac{4\sinh^2(n+\frac{1}{2})\alpha -(2n+1)^2\sinh^2\alpha}{2\sinh(2n+1)\alpha+(2n+1)\sinh2\alpha}\bigg]
\end{equation}
and
\begin{equation}
\lambda_\text{M} = -\sinh\alpha\sum_{n=1}^{\infty}\frac{2n(n+1)}{(2n-1)(2n+3)}\bigg[1 - \frac{4\cosh^2(n+\frac{1}{2})\alpha +(2n+1)^2\sinh^2\alpha}{2\sinh(2n+1)\alpha-(2n+1)\sinh2\alpha}\bigg],
\end{equation}
are series connected, respectively, to those appearing in Stimson and Jeffery \cite{Stimson1926} and Maude \cite{Maude1961}, with
\begin{equation}
\mathbb{M}_{1}^\text{AC} = 4\sinh^3\alpha\sum_{n=1}^{\infty}n(n+1)\bigg[\frac{1-\cosh(2n+1)\alpha+\sinh(2n+1)\alpha}{2\sinh(2n+1)\alpha+(2n+1)\sinh2\alpha}\bigg],
\end{equation}
%
%M1_BD
\begin{equation}
\mathbb{M}_{1}^\text{BD} = 4\sinh^3\alpha\sum_{n=1}^{\infty}n(n+1)\bigg[\frac{1+\cosh(2n+1)\alpha-\sinh(2n+1)\alpha}{2\sinh(2n+1)\alpha-(2n+1)\sinh2\alpha}\bigg], 
\end{equation}
\begin{align}
\mathbb{M}_{2}^\text{AC} &= \frac{16}{3}\sinh\alpha\sum_{n=1}^{\infty}\bigg\{\frac{n(n+1)}{(2n-1)(2n+3)}\bigg[\cosh(n-\frac{1}{2})\alpha - \cosh(n+\frac{3}{2})\alpha\bigg] \nonumber \\ &\times \frac{M_2}{2\sinh(2n+1)\alpha+(2n+1)\sinh2\alpha}\bigg\},
\end{align}
and
\begin{align} \label{eq_M2_BD_series}
\mathbb{M}_{2}^\text{BD} &= \frac{16}{3}\sinh\alpha\sum_{n=1}^{\infty}\bigg\{\frac{n(n+1)}{(2n-1)(2n+3)}\bigg[\sinh(n-\frac{1}{2})\alpha - \sinh(n+\frac{3}{2})\alpha\bigg] \nonumber \\ &\times \frac{M_2}{2\sinh(2n+1)\alpha-(2n+1)\sinh2\alpha}\bigg\}\cdot
\end{align}
In the limit $\alpha\to\infty$, where the two squirmers are widely separated, the coefficients obtained from equations (\ref{eq_l_sjs})–(\ref{eq_M2_BD_series}) approach their isolated values: $\lambda_\text{SJ} \rightarrow -3/2, \lambda_\text{M} \rightarrow 3/2$, $\mathbb{M}_1^\text{AC} \rightarrow 1$,  $\mathbb{M}_1^\text{BD} \rightarrow 1$, and $\mathbb{M}_2^\text{AC} = \mathbb{M}_2^\text{BD} \rightarrow 0$. Consequently, equations (\ref{eq_U1}) and (\ref{eq_U2}) reduce to the expected isolated swimming speeds, $U_{\alpha, \beta} = \frac{2}{3}B_1^{\alpha, \beta}$, as they should in the limit of vanishing hydrodynamic interactions.

\section{Numerical Simulations}\label{sec:Numerical}
To complement the theoretical analysis in \S \ref{sec:TheoreticalAnalysis}, numerical computations of the momentum and continuity equations are also performed using the finite element method implemented in COMSOL Multiphysics. The simulation is set up as a two-dimensional axisymmetric domain, with the two squirmers positioned along the axis of symmetry. To approximate an unbounded fluid domain, a sufficiently large computational region of $501a$ in width and $1014a$ in length is chosen to reduce confinement effects. The simulation domain is discretized using P3--P2 (third-order for fluid velocity and second-order for pressure) triangular mesh elements, with local mesh refinement applied around the squirmers. The total number of elements is on the order of 22,000 for Newtonian simulations and 50,000 for non-Newtonian simulations. All simulations are performed using the Multifrontal Massively Parallel Sparse (MUMPS) direct solver.

\begin{figure}[t]
    \centering\includegraphics[width=0.9\textwidth]{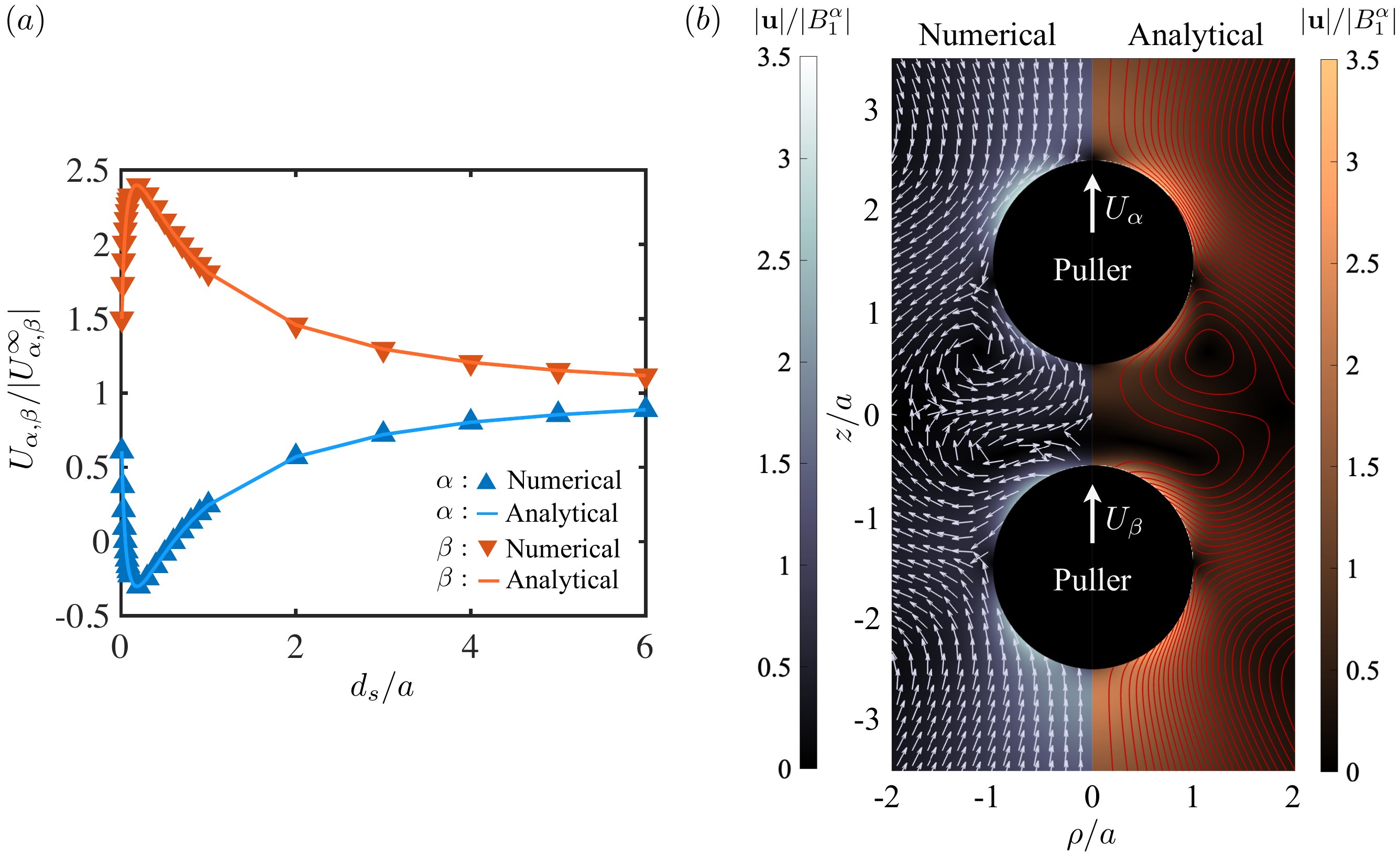}
    \caption{Comparisons between analytical and numerical solutions in terms of (a) the swimming speeds and (b) the flow field generated by a pair of identical puller-type squirmers with $B_1^\alpha = B_1^\beta>0$ and $[S^\alpha,S^\beta]=[5,5]$. In (a), analytical predictions (solid lines) are compared with numerical results (upward and downward triangles), as indicated in the legend, for varying scaled distances $d_s/a$. In (b), the color maps show the magnitude of the fluid velocity $|\mathbf{u}|$ scaled by the magnitude of the first squirming mode $|B_1^\alpha|$, for a pair of squirmers separated by a scaled distance of $d_s/a = 1$. The left half displays the numerical solution, with velocity directions indicated by arrows, whereas the right half displays the analytical solution, with streamlines shown as solid curves.}
    \label{fig:validation}
\end{figure}

The simulations are conducted in the laboratory reference frame, where the surface velocity distributions given by Eq.~(\ref{eqn:TwoModes}) together with the unknown swimming speeds $U_{\alpha,\beta}$ are prescribed on the squirmer surfaces as boundary conditions. The unknown swimming speeds, fluid velocity field, and pressure field are determined by solving the momentum and continuity equations simultaneously, with the force-free swimming condition for both squirmers enforced through global equations.  Our numerical implementation is validated against the exact solutions for single squirmers in a Newtonian fluid \cite{lighthill1952, blake1971}. In addition, we cross-validate the numerical simulations with the theoretical analysis presented in \S\ref{sec:TheoreticalAnalysis} by comparing the analytical predictions of the swimming speeds and the flow field with the corresponding numerical results, as shown in Fig.~\ref{fig:validation}. Fig.~\ref{fig:validation}(a) shows excellent agreement between the analytical solution (solid lines) and simulations (upward and downward triangles) for the swimming speeds. When the squirmers are far apart ($d_s/a \gg 1$), the swimming speeds approach those of isolated squirmers, \textit{i.e.,} $U_{\alpha,\beta} /|U^\infty_{\alpha,\beta}| \rightarrow 1$, as expected. Both analytical and numerical solutions capture the non-monotonic dependence of the swimming speeds when the squirmers are in close proximity to each other. The detailed features of the flow field are also in excellent agreement between the two methods, as shown in Fig.~\ref{fig:validation}(b). We discuss these results in more detail and then apply the validated framework to elucidate the swimming behaviors of squirmer pairs in various configurations in \S\ref{sec:Newtonian}.

In \S\ref{sec:Shear-thinning}, we further apply the numerical framework to explore regimes beyond the Newtonian limit by examining the effects of shear-thinning viscosity. Specifically, we employ the Carreau constitutive equation \cite{bird1987dynamics}, an inelastic, non-Newtonian constitutive model where the shear-rate-dependent viscosity in  the deviatoric stress $\bm{\tau}=\mu_s \dot{\bm{\gamma}}$ is given by 
\begin{align}
\mu_s = \mu_\infty+(\mu_0-\mu_\infty)\left(1+\lambda_C^2|\dot{\bm{\gamma}}|^2\right)^{\frac{n-1}{2}}. \label{eqn:Carreau}
\end{align}
Here, $\mu_0$ and $\mu_\infty$, respectively, represent the zero- and infinite-shear rate viscosities, $1/\lambda_C$ sets a characteristic strain rate at which the non-Newtonian feature starts to become significant, and $n$ is a power-law index determining the degree of shear thinning. This enables us to quantify how shear-thinning rheology modifies the swimming characteristics of a pair of squirmers. We first validate our numerical implementation by reproducing previous asymptotic and numerical results for an isolated squirmer in a shear-thinning fluid \cite{datt2015squirming}, before extending the simulations to examine pairwise interactions in  \S\ref{sec:Shear-thinning}.

\begin{figure}[t]
    \centering\includegraphics[width=1\textwidth]{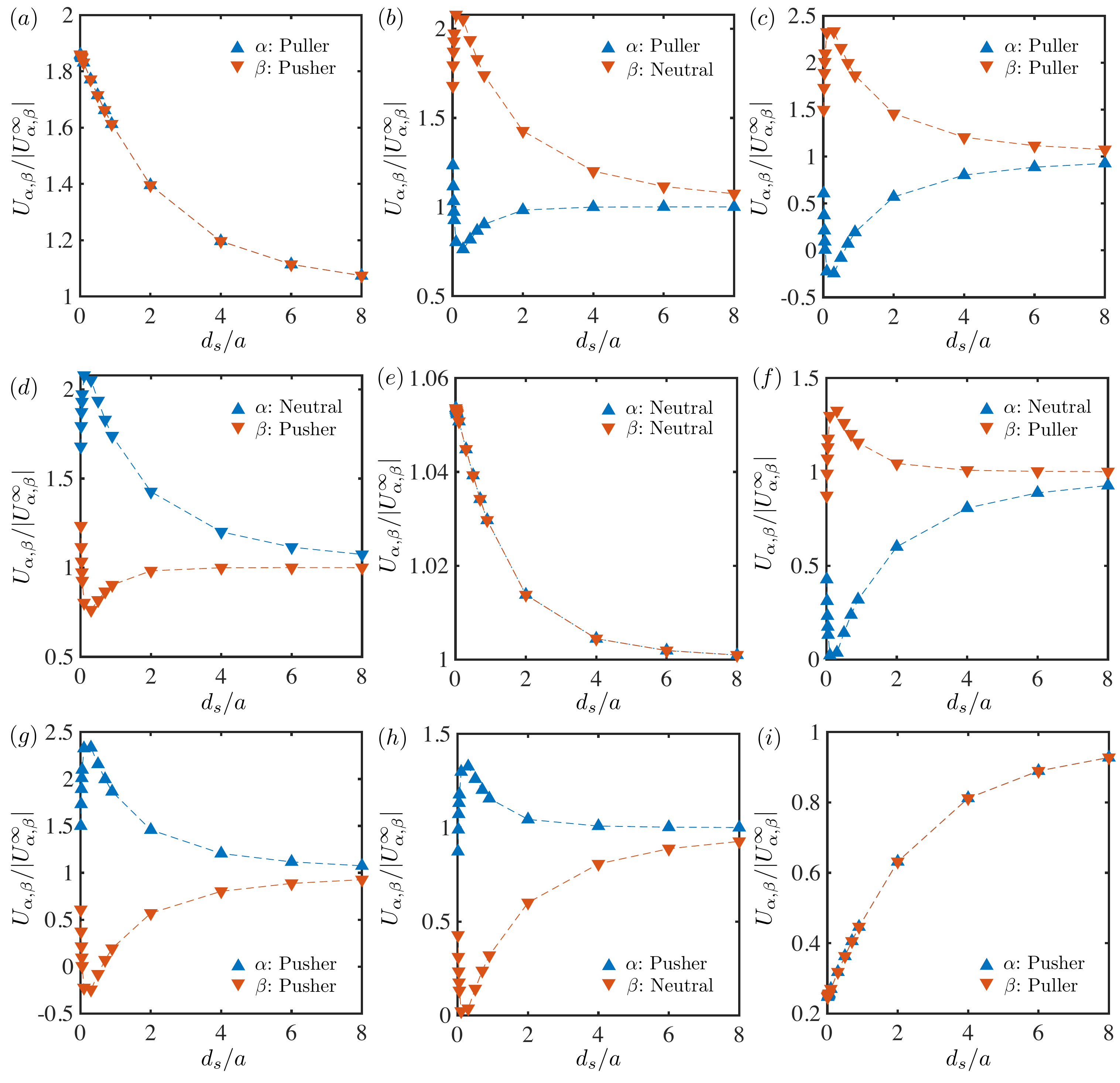}
    \caption{The swimming speeds of the squirmers, $U_{\alpha,\beta}$, scaled by the magnitudes of their corresponding isolated swimming speeds $U^\infty_{\alpha,\beta}$, are shown as functions of the scaled separation distance $d_s/a$ for: (a) puller–pusher with $[S^\alpha, S^\beta] = [5,-5]$; (b) puller–neutral with $[5,0]$; (c) puller–puller with $[5,5]$; (d) neutral–pusher with $[0,-5]$; (e) neutral–neutral with $[0,0]$; (f) neutral–puller with $[0,5]$; (g) pusher–pusher with $[-5,-5]$; (h) pusher–neutral with $[-5,0]$; and (i) pusher–puller with $[-5,5]$. Blue upward (orange downward) triangles denote the leading (trailing) squirmers.}
    \label{fig:SwimmingSpeed}
\end{figure}

\section{Results and Discussion} \label{sec:ResultsDiscussion}
We present results on the swimming characteristics of a pair of collinear squirmers under different configurations in this section. In the main text, we focus on configurations where both squirmers are prescribed with the same swimming modes ($B_1^\alpha = B_1^\beta>0$); that is, when isolated, both squirmers swim along the positive $z$--direction with the same speed. For these configurations, the scaled surface velocity distributions of squirmers $\alpha$ and $\beta$ are prescribed as $u_\theta^{\alpha,\beta}/B_1^{\alpha, \beta}=\sin \theta +S^{\alpha, \beta}\sin2\theta/2$, where $S^{\alpha, \beta} \equiv B_2^{\alpha,\beta}/B_1^{\alpha,\beta}$ is varied to examine the swimming behaviors associated with a puller ($S^{\alpha, \beta}>0$), a pusher ($S^{\alpha, \beta}<0$), and a neutral squirmer ($S^{\alpha, \beta}=0$). In particular, we identify co-swimming states, where both squirmers attain identical swimming speeds and travel together as a pair with a fixed separation. We analyze their propulsion performance in terms of both the speed and energetic cost of swimming. We also examine how the nonlinear effects introduced by shear-thinning rheology influence these co-swimming states in \S\ref{sec:Shear-thinning}. As a remark, head-on configurations do not give rise to co-swimming states; nevertheless, we include a discussion of these configurations in Appendix \ref{sec:Appendix} for completeness.

\subsection{Collinear swimming of a squirmer pair in a Newtonian fluid}\label{sec:Newtonian}

\begin{figure}[]
    \centering\includegraphics[width=1\textwidth]{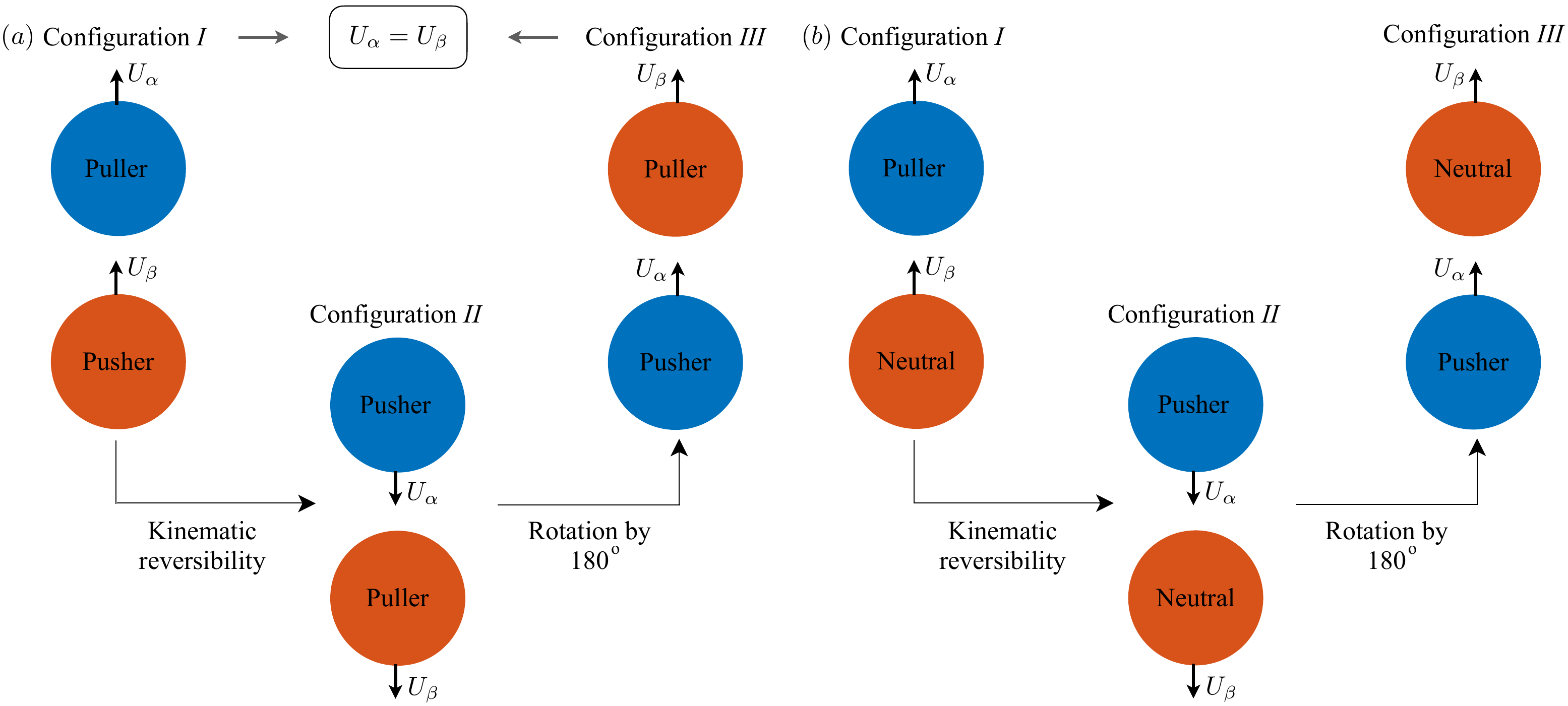} 
    \caption{Illustration of symmetry arguments for (a) puller–pusher and (b) puller–neutral configurations. Applying kinematic reversibility transforms configuration \textit{I} into configuration \textit{II}, and a subsequent 180° rotation yields configuration \textit{III}. In (a), configurations \textit{I} and \textit{III} are identical, leading to the emergence of a co-swimming state with $U_\alpha = U_\beta$. In (b), comparing configurations \textit{I} and \textit{III} elucidate the symmetries in swimming speeds observed in Figs.~\ref{fig:SwimmingSpeed}(b) and (d), where the speed of the front puller in Fig.~\ref{fig:SwimmingSpeed}(b) (configuration \textit{I}) corresponds to that of the rear pusher in Fig.~\ref{fig:SwimmingSpeed}(b) (configuration \textit{III}), while the neutral swimmer attains identical speeds in both configurations.}
    \label{fig:Symmetry}
\end{figure}

In this section, we examine the motion of two collinear squirmers and choose $S^{\alpha, \beta}  = 5, 0,$ and $-5$ to model a puller, a neutral squirmer, and a pusher, respectively, and examine their swimming behaviors under different pairings. Fig.~\ref{fig:SwimmingSpeed} displays several features of the hydrodynamic interaction between two squirmers, most notably the formation of co-swimming pairs in the puller-pusher [Fig.~\ref{fig:SwimmingSpeed}(a)], neutral-neutral [Fig.~\ref{fig:SwimmingSpeed}(e)], and pusher-puller [Fig.~\ref{fig:SwimmingSpeed}(i)] configurations. In these co-swimming pairs, the two squirmers translate with identical velocities for all separations. The emergence of these co-swimming configurations can be understood as a direct consequence of the kinematic reversibility of Stokes flows, as illustrated in Fig.~\ref{fig:Symmetry}(a) for a puller-pusher pair: applying kinematic reversibility transforms configuration \textit{I} into configuration \textit{II}, and a subsequent 180° rotation yields configuration \textit{III}. Since configurations \textit{I} and \textit{III} are identical in setup, it follows that the two squirmers must swim with equal velocities, $U_\alpha = U_\beta$, thereby establishing the co-swimming state.

\begin{figure}[]
    \centering\includegraphics[width=1\textwidth]{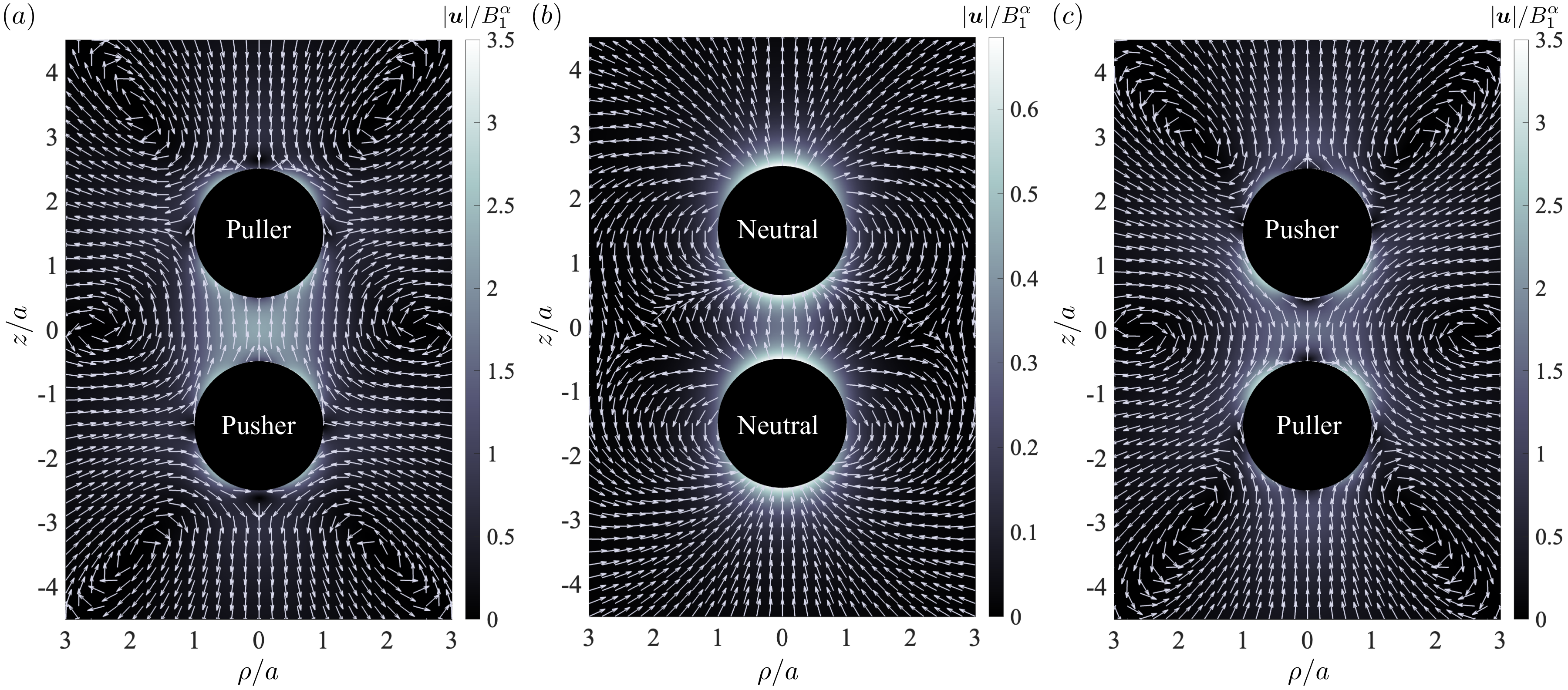} 
    \caption{Flow fields around the co-swimming pairs: (a) puller-pusher with $[S^\alpha, S^\beta]=[5, -5]$; (b) neutral-neutral with $[0,0]$; (c) pusher-puller with $[-5, 5]$. The color maps show the magnitude of the fluid velocity, and the arrows indicate the velocity directions. Here $d_s/a=1$ for all panels.} 
    \label{fig:FlowField}
\end{figure}

The magnitude of the co-swimming velocity depends strongly on the type of swimmers as well as their ordering. This dependence arises from the distinct hydrodynamic signatures of pushers and pullers. A pusher generates an extensile stresslet that drives fluid outward along the swimming axis, whereas a puller produces a contractile stresslet that draws fluid inward along the axis. In the puller–pusher configuration [Fig.~\ref{fig:SwimmingSpeed}(a)], the contractile stresslet of the leading puller and the extensile stresslet of the trailing pusher each induce a forward axial velocity on the other swimmer. These disturbances act constructively with their swimming motion, producing a marked increase in the co-swimming speed as the swimmers approach. In contrast, when the ordering is reversed [Fig.~\ref{fig:SwimmingSpeed}(i)], the extensile stresslet of the leading pusher and the contractile stresslet of the trailing puller induce axial velocities directed opposite to the swimming motion. The resulting hydrodynamic interaction diminishes the effective propulsion of both swimmers, leading to a pronounced reduction in the co-swimming speed. These features are reflected in the flow fields shown in Fig.~\ref{fig:FlowField}. When the puller leads and the pusher follows [Fig.~\ref{fig:FlowField}(a)], the axial flow in the inter-swimmer gap is aligned with the swimming direction, accompanying the observed increase in swimming speed. Reversing the ordering produces an opposing axial flow between the swimmers [Fig.~\ref{fig:FlowField}(c)], in conjunction with the significant slowdown in swimming speed.

The ordering-dependent enhancement and reduction of swimming speed observed here are qualitatively consistent with results reported for a dumbbell configuration in which two squirmers are connected by a short, dragless rigid rod \cite{Ishikawa2019}. In that mechanically constrained system, the two swimmers necessarily share the same propulsion speed due to the rigid linkage. By contrast, the present study considers freely interacting squirmers and shows that an identical co-swimming speed can emerge purely from hydrodynamic interactions, without any mechanical constraint. This observation is consistent with the previous study \cite{Ishikawa2019}, in which the net force on the rigid linkage vanishes for certain pairings of $S^\alpha$ and $S^\beta$. Moreover, this co-swimming behavior persists over a wide range of swimmer separations, with both enhancement and reduction becoming increasingly pronounced when the swimmers are closer to each other. Finally, the neutral–neutral case [Figs.~\ref{fig:SwimmingSpeed}(e) and \ref{fig:FlowField}(b)], which lacks a stresslet contribution, exhibits only a modest increase in swimming speed, even at small separations.

%%%%% Now onto non-co-swimming pairs
For the non-co-swimming configurations, the influence of the stresslet, whether from a puller or a pusher, on the other swimmer can be understood in a similar manner. For example, in Fig.~\ref{fig:SwimmingSpeed}(b), the leading puller generates a contractile disturbance that substantially enhances (by more than a factor of two) the speed of the trailing neutral squirmer. Likewise, in Fig.~\ref{fig:SwimmingSpeed}(d), the trailing pusher expels fluid along the swimming direction of the leading neutral squirmer, again producing more than a twofold increase in the neutral squirmer’s speed. We also note the symmetry between the configurations shown in Figs.~\ref{fig:SwimmingSpeed}(b) and \ref{fig:SwimmingSpeed}(d): the speed of the front puller in Fig.~\ref{fig:SwimmingSpeed}(b) corresponds to that of the rear pusher in Fig.~\ref{fig:SwimmingSpeed}(d), while the neutral swimmer, located at the rear in Fig.~\ref{fig:SwimmingSpeed}(b) and at the front in Fig.~\ref{fig:SwimmingSpeed}(d), attains identical speeds in both cases. Such symmetries are again a direct consequence of the kinematic reversibility of Stokes flows, as illustrated in Fig.~\ref{fig:Symmetry}(b) for the puller-neutral and neutral-pusher pairs. Similar correspondences are observed between the configurations in Figs.~\ref{fig:SwimmingSpeed}(c) and \ref{fig:SwimmingSpeed}(g), as well as between Figs.~\ref{fig:SwimmingSpeed}(f) and \ref{fig:SwimmingSpeed}(h).

\begin{figure}[t]
    \centering\includegraphics[width=0.65\textwidth]{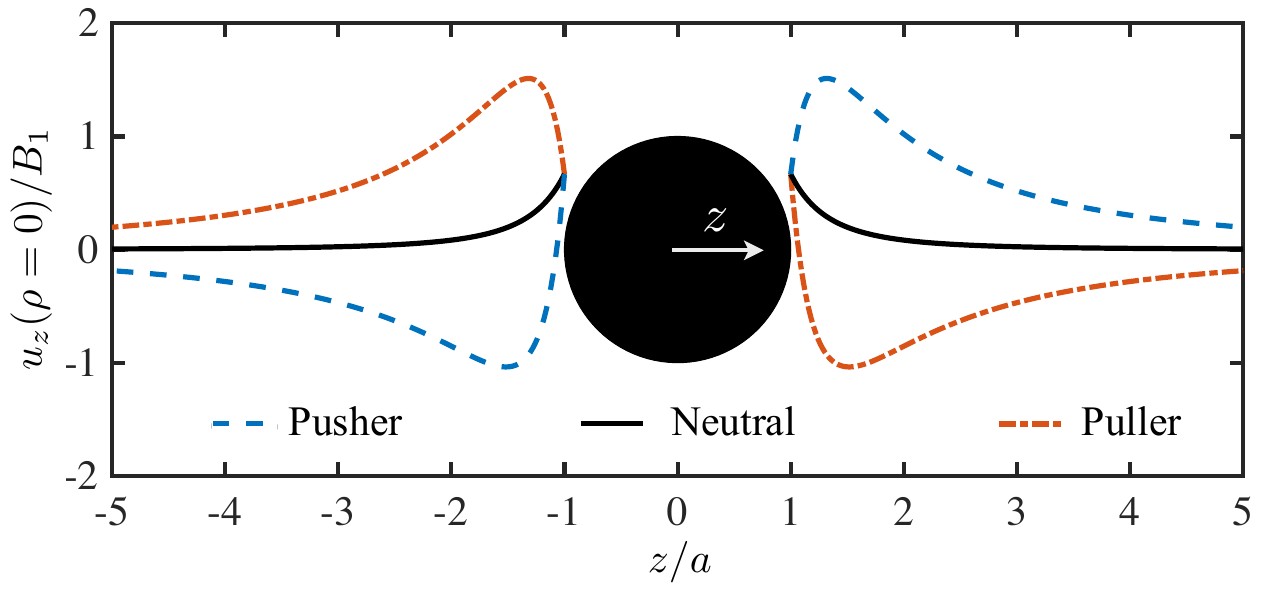}
    \caption{Axial flow velocity along the axis of symmetry ($\rho = 0$ or $\theta = 0$) as a function of the scaled axial distance $z/a$ from an isolated squirmer, for different values of the ratio $B_2/B_1$: a pusher with $B_2/B_1 = -5$ (blue dashed lines), a puller with $B_2/B_1 = 5$ (orange dot–dash lines), and a neutral squirmer with $B_2/B_1 = 0$ (black solid lines).}
    \label{fig:AxialFlow}
\end{figure}

We also remark that, in contrast to the monotonic dependence of the co-swimming speeds on separation, the swimming speeds of the non-co-swimming configurations exhibit non-monotonic variations as the inter-swimmer distance decreases. In these cases, local extrema in the swimmer velocities are observed at separations of $d_s/a \approx 0.2$. This behavior, consistently captured by both the exact and numerical solutions, highlights the significant role of near-field hydrodynamic interactions between the swimmers. This non-monotonic speed variation may be associated with the corresponding non-monotonic axial flow generated along the swimming direction of an isolated squirmer with the first $B_1$ and second $B_2$ squirming modes, as illustrated in Fig.~\ref{fig:AxialFlow}. In particular, the extrema of the axial flow  along the swimmer axis occur at $r/a = [\pm 1 + \sgn(B_2/B_1)\sqrt{1+8(B_2/B_1)^2}]/(2B_2/B_1)$, where the $+$ and $-$ signs apply to the front ($\theta =0$) and rear ($\theta = \pi$) axes, respectively. For a puller with $B_2/B_1 = 5$, the extrema are located at $r/a \approx 1.5 $ along the front axis and $r/a \approx 1.3 $ along the rear axis, which are of the same order as the separations at which local extrema in the swimmer velocities are observed. The positions are reversed for a pusher with $B_2/B_1=-5$. 

In particular, we examine the puller–puller configuration shown in Fig.~\ref{fig:SwimmingSpeed}(c). In this case, the leading puller enhances the swimming speed of the trailing puller through its contractile disturbance, while the trailing puller correspondingly reduces the speed of the leader. As a result, the two pullers approach one another. The opposite trend is observed for the pusher–pusher configuration shown in Fig.~\ref{fig:SwimmingSpeed}(g), where the extensile flows generated by each swimmer lead to mutual repulsion. These observations are consistent with earlier lattice Boltzmann simulations \cite{Llopis2010}. Our results further reveal pronounced non-monotonic variations as the swimmers enter the near-field regime. In this regime, strong hydrodynamic interactions can even cause one of the swimmers to reverse its direction of motion. This is evidenced by the negative velocities near the local extrema of the leading puller in Fig.~\ref{fig:SwimmingSpeed}(c) and the trailing pusher in Fig.~\ref{fig:SwimmingSpeed}(g).

\begin{figure}[t]
    \centering\includegraphics[width=1\textwidth]{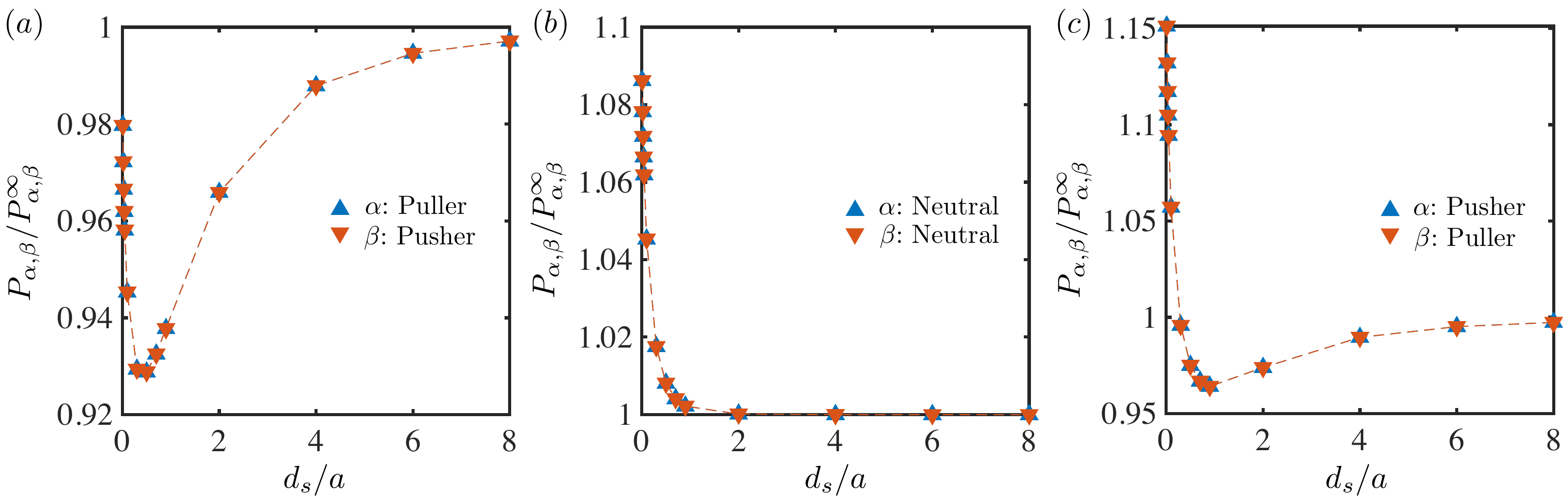}
    \caption{The power dissipation of the co-swimming squirmers, $P_{\alpha, \beta}$, scaled by their corresponding isolated power dissipation $P^\infty_{\alpha, \beta}$, are shown as functions of the scaled separation distance $d_s/a$: (a) puller-pusher with $[S^\alpha, S^\beta]=[5, -5]$; (b) neutral-neutral with $[0,0]$; (c) pusher-puller with $[-5, 5]$. Here, $d_s/a=1$ for all panels. Blue upward (orange downward) triangles denote the leading (trailing) squirmers.}
    \label{fig:Power}
\end{figure}

In addition to propulsion speed, the energetic cost of swimming provides another key metric for evaluating locomotion performance. The rate of work or power expended by each squirmer is computed as $P_{\alpha, \beta} = - \int_{A_{\alpha, \beta}} \boldsymbol{\sigma} \cdot \mathbf{n} \cdot \mathbf{u} \ \text{d}A$, where $\mathbf{n}$ is the unit outward normal on the squirmer surface $A_{\alpha, \beta}$. Focusing on the co-swimming pairs identified in Figs.~\ref{fig:SwimmingSpeed}(a), (e), and (i), the corresponding power dissipation is shown in Figs.~\ref{fig:Power}(a)–(c), respectively. Within each co-swimming configuration, the two squirmers attain not only identical propulsion speeds but also equal power dissipation. For the configuration consisting of a leading puller and a trailing pusher, the pair achieves a substantially enhanced propulsion speed [Fig.~\ref{fig:SwimmingSpeed}(a)] while simultaneously reducing power dissipation compared to swimming alone [Fig.~\ref{fig:Power}(a)]. In contrast, when the ordering is reversed (leading pusher and trailing puller), the pair swims slower [Fig.~\ref{fig:SwimmingSpeed}(i)] and expends more energy [Fig.~\ref{fig:Power}(c)] than the isolated case. For a neutral–neutral co-swimming pair, the modest enhancement in speed [Fig.~\ref{fig:SwimmingSpeed}(e)] is accompanied by a similarly modest increase in energy expenditure [Fig.~\ref{fig:Power}(b)].

\subsection{Effects of shear-thinning rheology on co-swimming pairs}\label{sec:Shear-thinning}

\begin{figure}[t]
    \centering\includegraphics[width=1\textwidth]{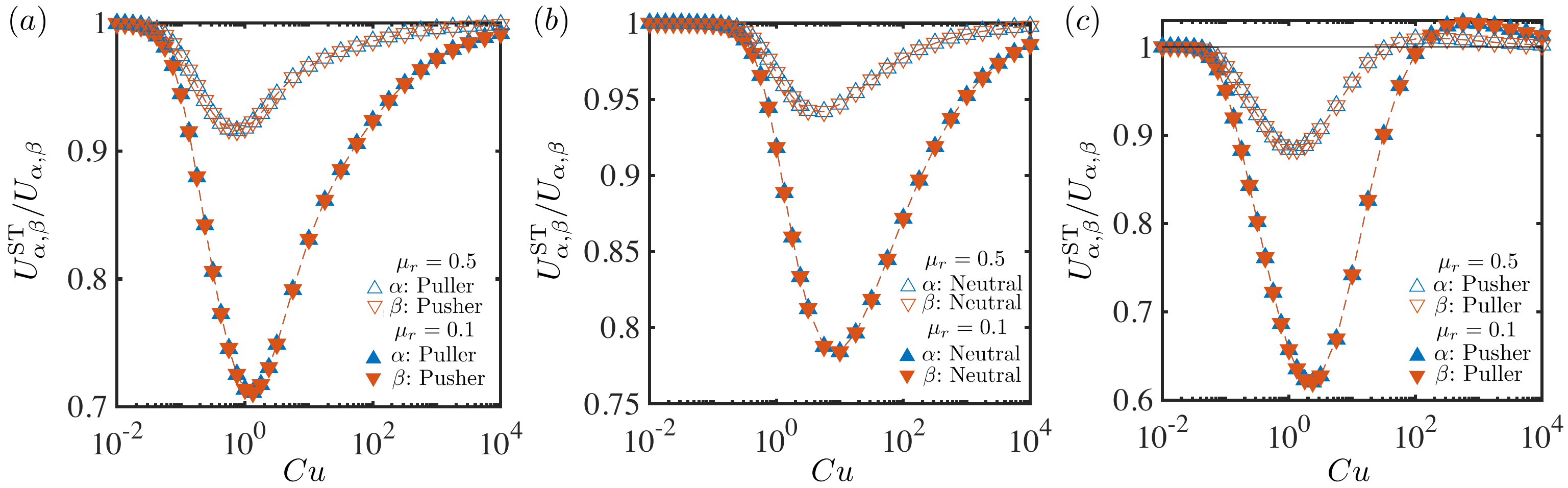}
    \caption{The swimming speed of co-swimming squirmers in a shear-thinning fluid, $U^\text{ST}_{\alpha,\beta}$, scaled by the corresponding Newtonian speed $U_{\alpha,\beta}$, are shown as functions of the Carreau number $Cu$ for viscosity ratios $\mu_r = 0.1$ (filled triangles) and $0.5$ (open triangles): (a) puller-pusher with $[S^\alpha, S^\beta]=[5, -5]$; (b) neutral-neutral with $[0,0]$; (c) pusher-puller with $[-5, 5]$. Here, $d_s/a=1$ and $n=0.25$ for all panels. Blue upward (orange downward) triangles denote the leading (trailing) squirmers.}
    \label{fig:STSwimmingSpeed}
\end{figure}

We now examine how shear-thinning rheology described by the Carreau constitutive equation [Eq.~(\ref{eqn:Carreau})] modifies the locomotion of the co-swimming states identified in the Newtonian limit [Figs.~3(a), (e), and (i)]. The swimming speeds in a shear-thinning fluid $U_{\alpha,\beta}^{ST}$ scaled by their corresponding Newtonian values $U_{\alpha, \beta}$ are shown in Fig.~\ref{fig:STSwimmingSpeed} for two viscosity ratios, $\mu_r = 0.5, 0.1$, over a wide range of Carreau number $Cu$. The Carreau number, $Cu = \omega \lambda_C$, compares the characteristic strain rate $\omega = |B_1^\alpha|/a$ to the crossover strain rate $1/\lambda_C$ at which non-Newtonian effects become significant, while the viscosity ratio $\mu_r = \mu_\infty/\mu_0$ sets the viscosity contrast between the zero- and infinite-shear-rate limits. For a fixed $n$, $Cu$ and $\mu_r$ determine the extent to which shear thinning manifests in the present flow.

\begin{figure}[t]
    \centering\includegraphics[width=1\textwidth]{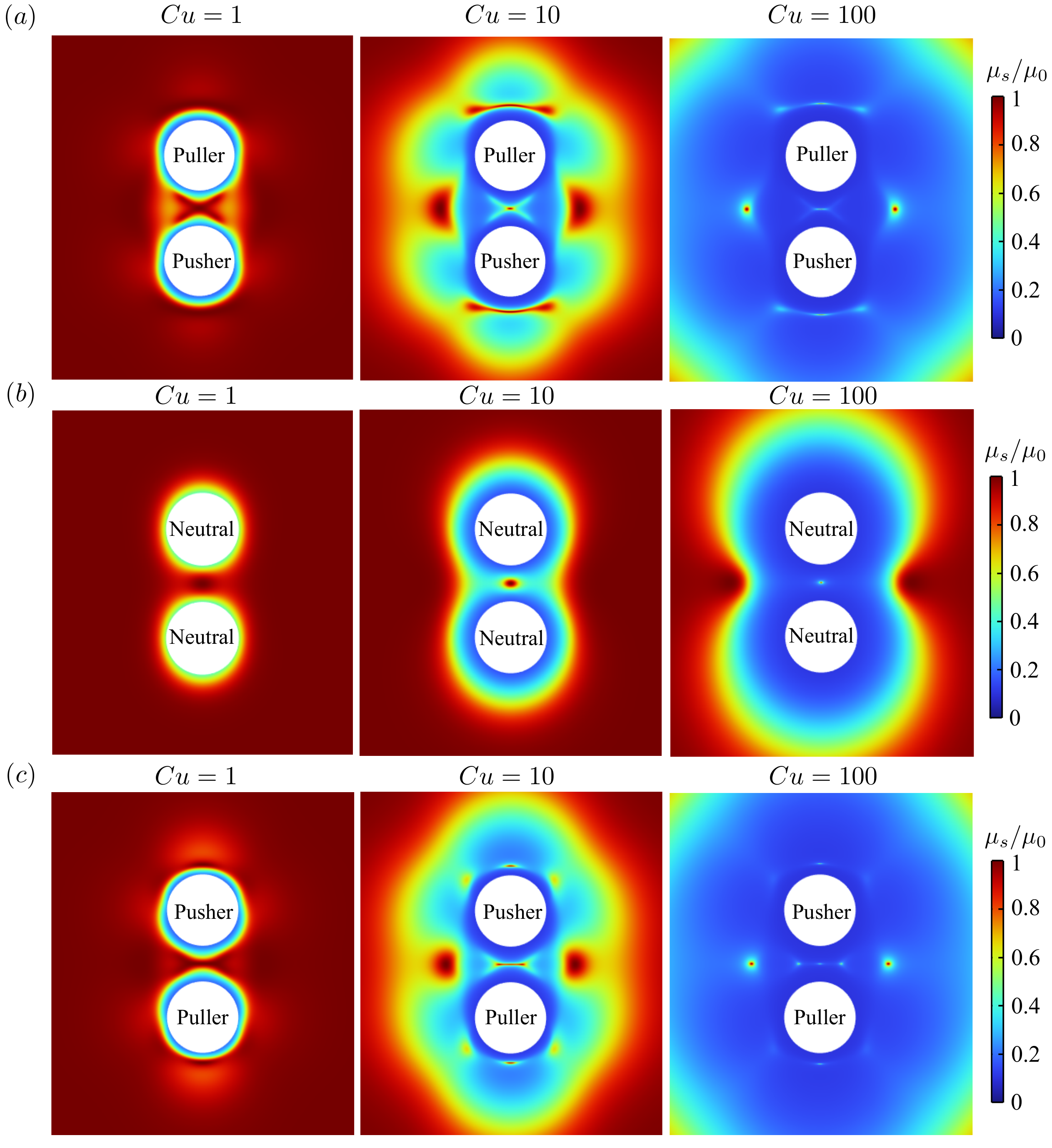}
    \caption{The viscosity map around the co-swimming squirmers at different Carreau number $Cu$: (a) puller-pusher with $[S^\alpha, S^\beta]=[5, -5]$; (b) neutral-neutral with $[0,0]$; (c) pusher-puller with $[-5, 5]$. Here, $d_s/a=1$, $\mu_r=0.1$, and $n=0.25$  for all panels. The color maps show the scaled viscosity.}
    \label{fig:ViscosityMap}
\end{figure}

Although the Carreau constitutive relation renders the problem nonlinear, it preserves a velocity-reversal symmetry provided the magnitude of the shear rate remains unchanged under reversal. In other words, reversing only the sign of the prescribed boundary velocities still leads to a corresponding reversal of the velocity field. This symmetry is more restricted than the rate-independent kinematic reversibility of Stokes flow \cite{lauga2009}. Nevertheless, the retained reversal symmetry in the shear-thinning fluid is sufficient for the symmetry argument underlying the Newtonian co-swimming states illustrated in Fig.~\ref{fig:Symmetry} to remain valid. Consequently, in all three configurations, the two squirmers continue to attain identical propulsion speeds in the shear-thinning fluid as shown in Fig.~\ref{fig:STSwimmingSpeed}. Shear-thinning rheology, therefore, modifies the magnitude of the co-swimming speed, but does not destroy the co-swimming state. In terms of the magnitude of the co-swimming speed, Fig.~\ref{fig:STSwimmingSpeed} reveals a non-monotonic dependence of speed on the Carreau number $Cu$. At small $Cu \ll 1$, the fluid behaves nearly Newtonian, and the swimming speed approaches its Newtonian value. As $Cu$ increases from this regime, all three co-swimming configurations exhibit a reduction in propulsion speed, reaching a local minimum at $Cu = O(1)\text{--}O(10)$. This reduction is more pronounced for smaller viscosity ratios, where shear-thinning effects are stronger. For $\mu_r = 0.1$, the speed decreases by approximately 20\%–40\% relative to the Newtonian co-swimming value. This slowdown reflects the heterogeneous viscosity distribution around the squirmers shown in Fig.~\ref{fig:ViscosityMap}, where low-viscosity regions form around zones of high shear. The resulting non-uniform stress distribution alters the hydrodynamic interactions between the two squirmers in a manner that ultimately diminishes their collective propulsion in this regime. As $Cu$ increases further beyond the minimum, the propulsion speed recovers and asymptotically approaches the Newtonian value as $Cu \to \infty$. These trends are largely consistent with prior observations for isolated squirmers in shear-thinning fluids \cite{datt2015squirming}, where propulsion is reduced relative to the Newtonian case for all two-mode squirmers. However, we remark on a modest speed enhancement for the pusher–puller configuration [Fig.~\ref{fig:STSwimmingSpeed}(c)], which is not observed for the puller–pusher [Fig.~\ref{fig:STSwimmingSpeed}(a)] or neutral–neutral [Fig.~\ref{fig:STSwimmingSpeed}(b)] pairs. Although small in magnitude, this enhancement highlights that shear-thinning rheology can either hinder or enhance co-swimming, depending on the detailed viscosity and stress distributions associated with specific swimmer configurations.

\begin{figure}[t]
    \centering\includegraphics[width=1\textwidth]{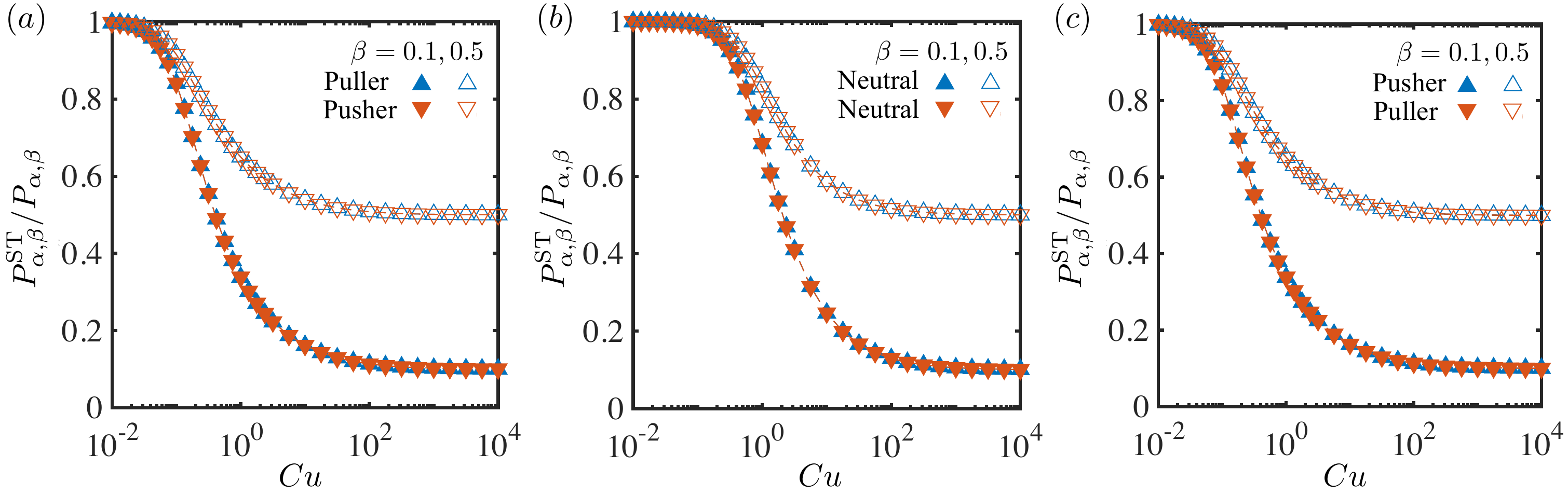}
    \caption{The power dissipation of co-swimming squirmers in a shear-thinning fluid, $P^\text{ST}_{\alpha,\beta}$, scaled by the corresponding Newtonian power dissipation $P_{\alpha,\beta}$, are shown as functions of the Carreau number $Cu$ for viscosity ratios $\mu_r = 0.1$ (filled triangles) and $0.5$ (open triangles): (a) puller-pusher with $[S^\alpha, S^\beta]=[5, -5]$; (b) neutral-neutral with $[0,0]$; (c) pusher-puller with $[-5, 5]$. Here, $d_s/a=1$ and $n=0.25$ for all panels. Blue upward (orange downward) triangles denote the leading (trailing) squirmers.}
    \label{fig:STPower}
\end{figure}

We also examine the energetic cost of swimming in a shear-thinning fluid. The power expended by each squirmer in a shear-thinning fluid $P^\text{ST}_{\alpha, \beta}$, normalized by its corresponding Newtonian value $P_{\alpha, \beta}$ (at $Cu=0$) is shown in Fig.~\ref{fig:STPower} for the three co-swimming configurations examined in Fig.~\ref{fig:STSwimmingSpeed}. Consistent with the symmetry of the co-swimming states, the two squirmers in each pair expend identical power across the full range of $Cu$, regardless of whether a swimmer is leading or trailing. This equality arises from the velocity-reversal symmetry preserved by the Carreau constitutive relation. Fig.~\ref{fig:STPower} also reveals that shear-thinning rheology consistently reduces the energetic cost of swimming relative to the Newtonian case. For all configurations and viscosity ratios examined, $P^\text{ST}_{\alpha, \beta}/P_{\alpha, \beta} \le 1$ over the entire range of $Cu$ shown in Fig.~\ref{fig:STPower}. The reduction in power becomes increasingly pronounced as $Cu$ grows and as the viscosity ratio decreases. For the stronger shear-thinning case ($\mu_r=0.1$), the power expenditure can drop to around 10\% of the Newtonian value at large $Cu$. This monotonic decrease in energetic cost contrasts with the non-monotonic behavior observed for the swimming speed in Fig.~\ref{fig:STSwimmingSpeed}. While the propulsion speed may either increase or decrease depending on $Cu$, the energetic cost is always reduced in a shear-thinning fluid, mirroring the decrease of viscosity with increasing shear rate described by the Carreau constitutive relation [Eq.~(\ref{eqn:Carreau})]. Physically, this behavior reflects that the dominant mechanism governing the energetic cost is the local reduction of viscosity in regions of high shear rates surrounding the swimmer. As the viscosity diminishes in these regions, the viscous stresses required to sustain the swimming motion decrease, thereby lowering the mechanical power expended during locomotion.

\section{Conclusions}\label{sec:Conclusions}

In this work, we have presented an analytical and numerical investigation of the hydrodynamic interactions between a pair of squirmers swimming collinearly in Newtonian and shear-thinning fluids. Using bispherical coordinates, we derived an exact, closed-form solution for the Stokes flow generated by two interacting squirmers and validated the result against finite element simulations, finding excellent agreement across all configurations considered. The exact solution complements previous analyses based on the reciprocal theorem approach by providing direct access to the detailed flow field around the squirmers, while also serving as a benchmark for validating numerical simulations of locomotion problems. In the Newtonian limit, our results reveal that the swimming behavior of a squirmer pair is governed by a subtle interplay between swimmer type, relative ordering, and separation distance. We identified co-swimming configurations where the two free-swimming squirmers translate with identical speeds over a wide range of separations, emerging purely from hydrodynamic interactions without any mechanical constraint. Among these, the puller–pusher configuration exhibits a pronounced enhancement in swimming speed at small separations while simultaneously reducing the energetic cost of propulsion, whereas the reversed pusher–puller ordering leads to a significant slowdown and increased power expenditure. In contrast, non-co-swimming configurations display non-monotonic variations in swimming speed as the squirmers enter the near-field regime, including local extrema and even reversals of swimming direction, highlighting the importance of accounting for near-field hydrodynamic interactions between the swimmers.

Extending the analysis to shear-thinning fluids, we demonstrated that the co-swimming configurations identified in the Newtonian case persist despite the nonlinear rheology. While shear-thinning modifies the magnitude of the co-swimming speed, it does not break the symmetry underlying these states. While the dependence of the propulsion speed on the Carreau number is largely similar to that observed for an isolated two-mode squirmer in a shear-thinning fluid \cite{datt2015squirming}, which consistently swims more slowly than in a Newtonian fluid, interacting pairs exhibit additional complexity. In particular, although the co-swimming speed is generally reduced relative to the Newtonian case over a broad intermediate range of $Cu$, modest speed enhancement can arise in specific configurations, such as the puller–pusher pair. This behavior reflects the complex, spatially heterogeneous viscosity fields that arise from the coupled flow generated by the two swimmers. In terms of the energetic cost of swimming, in all co-swimming configurations, shear-thinning rheology consistently reduces the mechanical power expended by the swimmers, even in regimes where the swimming speed is enhanced. 

To conclude, our results establish quantitative benchmarks for interacting squirmers in both Newtonian and shear-thinning fluids, motivating several avenues for future research. A possible next step is to extend the present framework to many-body systems to determine how the pairwise interactions identified here shape collective dynamics in suspensions of microswimmers in complex fluids. In particular, the role of shear-thinning rheology in collective behavior remains relatively unexplored. Moreover, many biological fluids exhibit not only shear-thinning viscosity but also viscoelasticity. Probing how these rheological effects interact, whether synergistically or competitively, to influence cooperative swimming presents another direction for future study. In addition, while the present work focuses on equal-sized squirmers, the theoretical framework developed here readily accommodates swimmers of unequal sizes. Investigating how size asymmetry alters pairwise interactions represents a natural next step toward understanding the dynamics of heterogeneous suspensions in complex fluids. Finally, the viscosity gradients considered here are self-generated through swimmer-induced shear. By contrast, an area of growing interest concerns the tactic behavior of swimmers in externally imposed viscosity gradients \cite{Liebchen2018, Datt2019, Anand2025, Kobayashi_Yamamoto_2025}. Extending the present framework to examine pairwise interactions under externally prescribed viscosity gradients may provide further insight into the dynamics of active particles in different complex heterogeneous environments.

\section*{Acknowledgments}
Y.N.Y.~acknowledges support from  the  National Science Foundation (DMS-1951600 and DMS-2510714) and Flatiron Institute, part of Simons Foundation. O.S.P.~acknowledges partial support from  the National Science Foundation (CBET-2323046 and CBET-2419945). C.T.L.~acknowledges support from the National Science and Technology Council of Taiwan (114-2917-I-564-016) and the use of high-performance computing resources at NJIT and SCU.\\

\appendix

\begin{figure}[]
    \centering\includegraphics[width=1\textwidth]{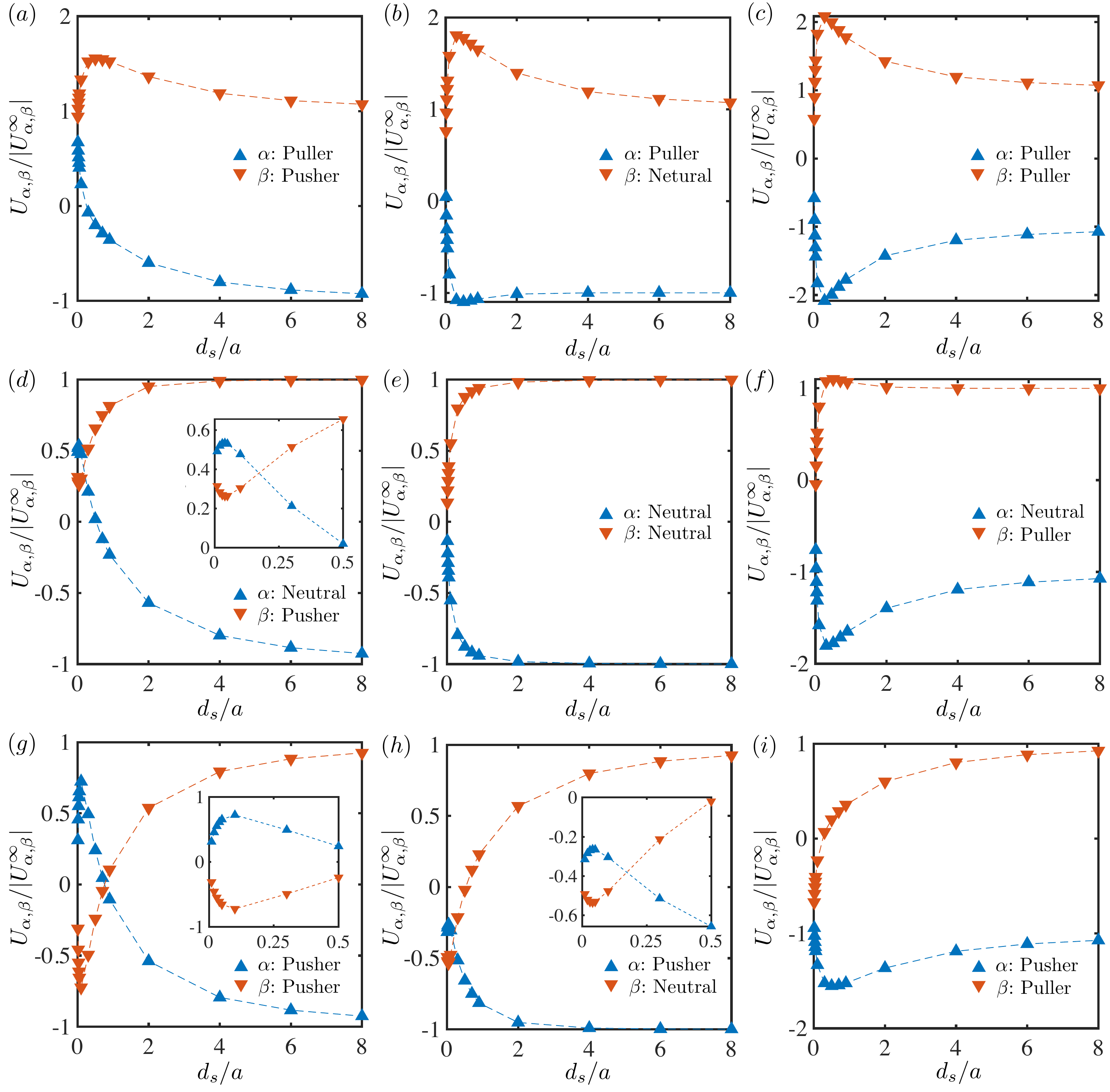} 
    \caption{The swimming speeds of the squirmers, $U_{\alpha,\beta}$, scaled by the magnitudes of their corresponding isolated swimming speeds $U^\infty_{\alpha,\beta}$, are shown as functions of the scaled separation distance $d_s/a$ for: (a) puller–pusher with $[S^\alpha, S^\beta] = [5,-5]$; (b) puller–neutral with $[5,0]$; (c) puller–puller with $[5,5]$; (d) neutral–pusher with $[0,-5]$; (e) neutral–neutral with $[0,0]$; (f) neutral–puller with $[0,5]$; (g) pusher–pusher with $[-5,-5]$; (h) pusher–neutral with $[-5,0]$; and (i) pusher–puller with $[-5,5]$. Blue upward (orange downward) triangles denote the leading (trailing) squirmers.} 
    \label{fig:HHSwimming}
\end{figure}

\section{Head-on configurations}\label{sec:Appendix}

In this Appendix, we examine the case of oppositely signed swimming modes ($B_1^\alpha = -B_1^\beta<0$), which corresponds to a head-on configuration, with the two squirmers swimming directly toward one another in the isolated limit. For these configurations, the scaled surface velocity distributions of squirmers $\alpha$ and $\beta$ are prescribed as $u^{\alpha,\beta}/B_1^\beta = \mp \sin\theta +S^{\alpha, \beta} \sin 2\theta/2$, where the minus sign is for squirmer $\alpha$ and the plus sign is for squirmer $\beta$. Across all pairings shown in Fig.~\ref{fig:HHSwimming}, in the far-field limit ($d_s/a \gg 1$), the two squirmers recover their isolated swimming speeds with opposite signs, namely $U_\alpha/U_\infty \rightarrow -1$ and $U_\beta/U_\infty \rightarrow 1$, swimming toward each other. When the squirmers are in closer proximity, the near-field interactions could cause different non-monotonic variations as a function of their separation distance.

We first note the antisymmetric swimming velocities in the puller–puller [Fig.~\ref{fig:HHSwimming}(c)], neutral–neutral [Fig.~\ref{fig:HHSwimming}(e)], and pusher–pusher [Fig.~\ref{fig:HHSwimming}(g)] configurations, in which geometric symmetry leads the two squirmers to swim with equal magnitudes but opposite signs of velocity. While the puller–puller and neutral–neutral pairs maintain the same swimming directions across all separation distances shown, at smaller separations the hydrodynamic repulsion between the pushers in Fig.~\ref{fig:HHSwimming}(g) slows their approach to a stagnant state ($U_\alpha=U_\beta=0$) and, upon further reduction in separation, causes a reversal of their swimming directions. 

When the geometric symmetry is broken, the two squirmers develop swimming velocities with different magnitudes and directions, as shown in Figs.~\ref{fig:HHSwimming}(a), (b), (d), (f), (h), and (i). Nevertheless, symmetries exist across different configurations due to kinematic reversibility, as observed in Figs~\ref{fig:HHSwimming}(a) \& (i), (b) \& (f), (d) \& (h), where the ordering of the squirmers is exchanged, together with reversals in the signs of the swimming velocities. In these configurations, although the two squirmers swim toward each other when they are far apart, near-field interactions at sufficiently small separations can cause them to translate in the same direction, highlighting the importance of accurately resolving near-field hydrodynamic interactions to capture qualitatively correct behavior.

\bibliography{sample}

\end{document}